\documentclass[preprint,1p,sort&compress]{elsarticle}
\usepackage{bm}
\usepackage{graphics}

\newcommand{\be}{\begin{equation}}
\newcommand{\ee}{\end{equation}}
\newcommand{\bea}{\begin{eqnarray}}
\newcommand{\eea}{\end{eqnarray}}
\newcommand{\BM}{\begin{bmatrix}}
\newcommand{\EM}{\end{bmatrix}}

\newcommand{\ve}{\varepsilon}
\newcommand{\bra}[1]{\bigl\langle #1 \bigr|}
\newcommand{\ket}[1]{\bigl| #1 \bigr\rangle}
\newcommand{\inttt}{\int_{t_i}^{t_f}\!\!}
\newcommand{\brasub}{{\!\!\phantom{\big\langle}}}
\newcommand{\PB}[2]{\bm{\{}\!\{ #1 ,\, #2 \}\!{\bm \}}}

\journal{Physics Letters A}
\begin{document}

\begin{frontmatter}

\title{From Classical Mechanics with Doubled Degrees of Freedom 
to Quantum Field Theory for 
Nonconservative System}

\author{Y.~Kuwahara\corref{cor}}
\ead{a.kuwahara1224@asagi.waseda.jp}

\author{Y.~Nakamura}
\ead{nakamura@aoni.waseda.jp}

\author{Y.~Yamanaka}
\ead{yamanaka@waseda.jp}

\cortext[cor]{Corresponding author. Tel.: +81-3-5286-8092}

\address{Department of Electronic and Photonic Systems, Waseda University, Tokyo 169-8555, Japan}

\begin{abstract}
The $2 \times 2$-matrix structure of Green's functions is a common feature for the real-time formalisms of quantum field theory under thermal situations, such as the closed time path formalism and Thermo Field Dynamics (TFD). It has been believed to originate from quantum nature. Recently, Galley has proposed the Hamilton's principle with initial data for nonconservative  classical systems, doubling each degree of freedom [C.~R.~Galley, 2013].
We show that the Galley's Hamilton formalism can be extended to quantum field and that the resulting theory is naturally identical with nonequilibrium TFD.
\end{abstract}


\end{frontmatter}





\section{Introduction}
The essence of the formalism of classical nonconservative systems
by Galley \cite{Galley} is the doubling of degree of freedom.
As is well known, the use of $2 \times 2$-matrix Green's functions
is common for the real-time descriptions of 
a quantum system under thermal situations.  The typical examples are
the closed time path (CTP) formalism, also called the Keldysh-Schwinger
 formalism \cite{CTP}, and Thermo Field Dynamics (TFD) \cite{TU, UMT, AIP}. 

We first note that the $2 \times 2$-matrix structure does not necessarily
mean the doubling of degree of freedom.
In CTP, the origin of the $2 \times 2$-matrix structure is the doubling
of time-paths, i.e.\,forward and backward time-paths, which are introduced 
to evaluate the quantum expectation value of time-dependent operators.
TFD, which had been formulated for equilibrium \cite{TU,UMT} 
and later extended to
nonequlibrium \cite{AIP}, starts with the doubling of every degree of freedom, 
so that the thermal average by a density matrix is replaced with that 
of a pure state, called the thermal vacuum. The decisive point is that 
the operators on the forward time-path and backward one in CTP do not 
(anti)-commute with each other in general, while a pair of the doubled operators
in TFD are independent canonical variables and (anti)-commute with each other.
Although CTP and TFD are similar superficially when the 
$2 \times 2$-matrix Green's functions are treated, both the formalisms
should be recognized as different ones.

The Galley's formalism for classical nonconservative mechanics, given in 
both the Lagrange and Hamilton formalisms \cite{Galley},
indicates a new and profound aspect of doubling of degree of freedom, 
namely it is not simply of quantum origin but universal for all the dynamics
from classical mechanics to quantum theory.
In this Letter we show that the Galley's classical Hamilton formalism
  can be extended to quantum field system and that the formalism 
thus obtained becomes the nonequilibrium TFD \cite{AIP, NY2011, NY2013}
quite naturally.

\section{Classical formalism}
According to the {\em Hamilton's principle with initial data\/} \cite{Galley}, 
the Lagrange formalism is given as follows.  Consider a system described
by a set of generalized
coordinates and velocities, $q={q^j}$ and ${\dot q}={{\dot q}^j}$
 $(j=1,\cdots, N)$, and each of them is doubled as $q \rightarrow (q_1,q_2)$
and ${\dot q} \rightarrow ({\dot q}_1,{\dot q}_2)$.  The action functional 
of $q_1$ and $q_2$ is defined as the line integrals of $L(q_1,{\dot q}_1)$
from $t_i$ to $t_f$ and of $L(q_2,{\dot q}_2)$ from $t_f$ to $t_i$ plus 
the line integral of a possible function $K$, depending on both of the 
variables and representing nonconservative forces,
\bea \label{eq:ClassicalAction}
S[q_\alpha] &=& \inttt \Lambda(q_\alpha,{\dot q}_\alpha,t)\, dt \,, \\
\Lambda(q_\alpha,{\dot q}_\alpha,t)&=&L(q_1,{\dot q}_1)-L(q_2,{\dot q}_2)
+ K(q_\alpha,{\dot q}_\alpha,t) \,,
\eea
where $q_\alpha$ stands for $(q_1,q_2)$. The function $K$ is absent
 before integrating out environmental variables, but appears after integrating out them.
The variations are taken with the fixed initial values, 
$q_\alpha(t_i)=q_{\alpha I}$ and
$\delta q_\alpha(t_i)=0$, $q_{1I}$ and $q_{2I}$ being independent of each other.
At $t=t_f$, we only require $q_{1}(t_f)=q_{2}(t_f)$ and
 ${\dot q}_{1}(t_f)={\dot q}_{2}(t_f)$, called the {\em equality 
condition\/} at $t=t_f$, but neither the fixed values of $q_\alpha (t_f)$
nor $\delta q_\alpha(t_f)=0$. After all the calculation, we take the
 {\em physical limit\/}, $q_1(t)=q_2(t)$ for all $t$, which includes
 $q_{1I}=q_{2I}$.

The corresponding canonical formalism can be constructed. 
One can start with $K=0$ in quantum field model, since
the dissipative property of a system 
 appears through interactions with environmental variables of an 
infinite degrees of freedom such as 
the reservoir variables. For convenience,
the metric factor $\ve_\alpha$ is introduced as 
$
	\ve_1 = 1 \,, \ve_2 = -1 \,.
$
The generalized momenta are defined by
$
	p_\alpha= \ve_\alpha \partial \Lambda/ \partial {\dot q}_\alpha=
	 \ve_\alpha {\partial L_\alpha}/{\partial {\dot q}_\alpha}
	 \,,
$
where $L_\alpha$ is $L(q_\alpha,{\dot q}_\alpha)$.
Note the presence of $\ve_\alpha$, which makes the formulations below simpler.
The stationary condition for the action reads
\be
	\delta S = \inttt \left\{ 
	\left(\frac{\partial L_1}{\partial q_1}-\frac{d p_1}{dt}\right)
	\delta q_1
	-\left(\frac{\partial L_2}{\partial q_2}-\frac{d p_2}{dt}\right)
	\delta q_2 \right\}\, dt +
	\left[p_1\delta q_1-p_2 \delta q_2 \right]_{t_i}^{t_f}=0 \, .
\ee
The surface term vanishes because of $\delta q_1(t_i)=\delta q_2(t_i)=0$
and the equality condition, implying 
$ p_1(t_f)=p_2(t_f)$ and $\delta q_1(t_f)=\delta q_2(t_f)$.
The Hamiltonian is
\be	\label{eq:H}
	H= \sum_{\alpha=1}^2 \ve_\alpha \left(p_\alpha{\dot q}_\alpha -L_\alpha\right) 
	=\sum_{\alpha=1}^2 \ve_\alpha H_\alpha =H_1-H_2\, ,
\ee
and the canonical equations are
\be
	{\dot q}_\alpha = \ve_\alpha \frac{\partial H}{\partial p_\alpha} 
	\, , \qquad {\dot p}_\alpha = -\ve_\alpha 
	\frac{\partial H}{\partial q_\alpha} \, .
	\label{eq:canonicaleq}
\ee
According to Ref.~\cite{Galley}, one can introduce 
the Poisson bracket with the metric,
\be
	\PB{A}{B} = \sum_{\alpha,j} \ve_\alpha  \left( 
	 \frac{\partial A}{\partial q^j_\alpha}\frac{\partial B}{\partial p^j_\alpha}
	-\frac{\partial A}{\partial p^j_\alpha}\frac{\partial B}{\partial q^j_\alpha}
	\right) \, .
\ee

\section{Canonical quantization}
The canonical commutation relations are 
$
	\PB{q^j_\alpha}{p^{j'}_{\alpha'}} = \ve_\alpha \delta_{\alpha\alpha'}\delta^{jj'}
	\,,\;\mathrm{otherwise}=0 \,.
$
The canonical equations (\ref{eq:canonicaleq}) are rewritten
\be \label{clHaEq}
	\dot{q}_\alpha = \PB{q_\alpha}{H} \,,\qquad
	\dot{p}_\alpha = \PB{p_\alpha}{H} \, .
\ee

The canonical quantization of 
the classical formalism with doubled degrees
of freedom above yields its
quantum version in the Heisenberg picture, namely the canonical variables
are interpreted as the Heisenberg operators and the Poisson bracket 
is replaced with the commutation relation (for simplicity only bosons are
considered):
$
	[{q^{j}_{\alpha H}}(t)\,,p^{j'}_{\alpha'H}(t)] =i
	 \ve_\alpha \delta_{\alpha\alpha'}\delta^{jj'}
	\,
$ and so on.
The canonical equations (\ref{clHaEq}) are transformed into
the Heisenberg equations,
$
	 i \dot{q}_{\alpha H}(t) = [q_{\alpha H}(t) , H] \,,\;
	 i \dot{p}_{\alpha H}(t) = [p_{\alpha H}(t) , H] \,.
$
One may use the operators $A_\pm=(A_1 \pm A_2)/\sqrt{2}\,,$ 
instead of $A_1,A_2$, as was done in Ref.~\cite{Galley},
but they are not used in our approach.

The next step is to construct the bra- and ket-states,
denoted by $\bra{\Psi_b}$ and $\ket{\Psi_k}$, 
whose matrix elements of the Heisenberg operators, 
$\bra{\Psi_b} \mathrm{\{Heisenberg operators\}} \ket{\Psi_k}$,
should be c-number observable values and reproduce the classical
ones in the classical limit.  It is not necessary in general that
$\bra{\Psi_b}$ is an Hermitian conjugate of $\ket{\Psi_k}$. 
The key point in this Letter is how to transfer 
the equality condition and the physical limit in classical theory
to quantum theory. The conditions should be put on 
$\bra{\Psi_b}$ and $\ket{\Psi_k}$, 
since it is not allowed to express them as the operator equalities.
First we introduce the exchange operation of the operators, 
$q_{1H}(t) \leftrightarrow q_{2H}(t)$ and so on,
and denote it by ${}^\sim$, 
\be	
	\bigl[ q_{1H}(t) \bigr]^\sim = q_{2H}(t) \,,\qquad
	\bigl[ q_{2H}(t) \bigr]^\sim = q_{1H}(t) \,,
\ee
and similarly for $p_{1H}(t)$ and $p_{2H}(t)$.
The operation is defined to keep the operator ordering
and to commute with the operation of the Hermitian conjugate,
\bea
	\bigl(A_H(t_1)B_H(t_2)\bigl)^\sim &=& \tilde{A}_H(t_1) \tilde{B}_H(t_2) \,,\\
	\bigl(A^\dagger_H(t) \bigl) ^\sim &=&\bigl( {\tilde A}_H(t)\bigl)^\dagger \,.
\eea
From the canonical commutation relations, any c-number $c$ must
transform as
$
	c^\sim = c^\ast
$,
and furthermore
\be	
	\bigl(c_1 A_H(t) + c_2B_H(t))^\sim 
	= c_1^* \tilde{A}_H(t) + c_2^* \tilde{B}_H(t) \,.
\ee
The operation is consistent with the dynamics only when 
$
	H^\sim=-H
$.
This condition is fulfilled for the Hamiltonian in Eq.~(\ref{eq:H}).
It is remarked that the above operation is identical with the tilde-conjugation
in TFD \cite{AIP}, and the above rules are called the tilde-conjugation ones.

Let us put only two requirements below for the physical limit and 
the equality condition, leading to a consistent construction
of the states.
\begin{itemize}
\item[(a)] The physical limit is realized for the T-product matrix elements 
in the quantum theory as
\be
	\left(\bra{\Psi_b}{\rm T}[q_{1H}(t_1)p_{1H}(t_2)\cdots ]
	\ket{\Psi_k}\right)^\sim 
	= \bra{\Psi_b}{\rm T}[q_{2H}(t_1)p_{2H}(t_2)\cdots  ]\ket{\Psi_k} \, .
\ee
\item[(b)] The equality conditions are realized as the subsidiary conditions
 on the bra state,
\be	\label{eq:BraEquCond-1}
	\bra{\Psi_b}\left\{
	\begin{array}{c}
	q_{1H}(t_f)\\
	 p_{1H}(t_f)
	\end{array}
	\right\}=
	\bra{\Psi_b}\left\{
	\begin{array}{c}
	q_{2H}(t_f)\\
	 p_{2H}(t_f)
	\end{array}
	\right\} \, .
\ee
\end{itemize}

The requirement (a) immediately implies
\be	\label{eq:StatesTilde}
	\left(\bra{\Psi_b}\right)^\sim = \bra{\Psi_b} \qquad
	{\rm and}\qquad \left(\ket{\Psi_k}\right)^\sim=\ket{\Psi_k} \, .
\ee
The reason for taking the T-product is that it represents the microscopic
causality for quantum field system, and the usability of the
Feynman diagram method.

It can be shown from the requirement (b) that for any product of Hermitian
operators,
\be
	\bra{\Psi_b}A_{1H}(t_f)B_{1H}(t_f)\cdots 
	=\bra{\Psi_b}\cdots B_{2H}(t_f)A_{2H}(t_f) 
	=\bra{\Psi_b} \bigl(A_{2H}(t_f) B_{2H}(t_f)\cdots\bigr)^\dagger \, ,
\ee
and therefore that for the Hermitian $H_1$ and $H_2$
\be
	\bra{\Psi_b}H_1(t_f)= \bra{\Psi_b}H_2(t_f)\,,
     \quad \mathrm{i.e.,} \quad \bra{\Psi_b}H=0 \,.
\ee
This in turn implies that the subsidiary conditions at $t=t_f$ in 
Eq.~(\ref{eq:BraEquCond-1}) are extended for any $t$,
\be	\label{eq:BraEquCond-2}
	\bra{\Psi_b}\left\{
	\begin{array}{c}
	q_{1H}(t)\\
	 p_{1H}(t)
	\end{array}
	\right\}=
	\bra{\Psi_b}\left\{
	\begin{array}{c}
	q_{2H}(t)\\
	 p_{2H}(t)
	\end{array}
	\right\} \, .
\ee
We introduce a complete set $\{\ket{u_\ell}_1\}$
for the system of $(q_{1H}(t), p_{1H}(t))$,
and its time-reversed one, denoted by $\{\ket{{\bar u}_\ell}_2\}$,
for that of $(q_{2H}(t), p_{2H}(t))$. Note the relations,
\be
	\brasub_1\bra{u_\ell}\left\{
	\begin{array}{c}
	q_{1H}(t)\\
	 p_{1H}(t)
	\end{array}
	\right\}
	\ket{u_{\ell'}}_1 
	=\brasub_2 \bra{{\bar u}_{\ell'}}\left\{
	\begin{array}{c}
	q_{2H}(t)\\
	 p_{2H}(t)
	\end{array}
	\right\}
	\ket{{\bar u}_{\ell}}_2\, .	
\ee
Let us expand the state $\bra{\Psi_b}$ in terms of 
the complete set of the whole system $\brasub_1\bra{u_{\ell_1}} \otimes
\brasub_1\bra{{\bar u}_{\ell_2}}$, as
$
	\bra{\Psi_b}=\sum_{\ell_1,\ell_2} C_{\ell_1 \ell_2}
	\brasub_1\bra{u_{\ell_1}} \otimes
	\brasub_2\bra{{\bar u}_{\ell_2}}\,.\,
$ 
Then the conditions in Eq.~(\ref{eq:BraEquCond-2}) compel the matrix
 $C_{\ell_1 \ell_2}$ to be proportional to the unit matrix
except for a constant factor, 
\be	\label{eq:braI}
	\bra{\Psi_b}=\bra {I}\equiv \sum_{\ell} 
	\brasub_1\bra{u_{\ell}} \otimes
	\brasub_2\bra{{\bar u}_{\ell}}\, .
\ee

The requirements (a) and (b) are correlated. If the anti-T-product is chosen,
the subsidiary conditions must be those on the ket state.  Furthermore, when
one formulated the quantum version using  a general product of the operators,  
the subsidiary conditions both on the bra- and ket-states would be necessary.
It also guarantees from (a) and (b) that the expectation value of an Hermitian
 operator $A_1$, $\langle A (t)\rangle= \bra{\Psi_b}A_{1H}(t)\ket{\Psi_k}$,
is real.

\section{Interaction picture}
The general discussions of the quantum theory in the Heisenberg picture 
fix the bra-state as in Eq.~(\ref{eq:braI}), but the ket state $\ket{\Psi_k}$
only with the constraint in Eq.~(\ref{eq:StatesTilde}) is not determined yet.
In order to specify $\ket{\Psi_k}$, we move to the interaction picture,
because our main interest is in quantum field systems, which include 
systems attached with environmental ones with infinite degrees of freedom
such as a reservoir.
In quantum field theory, the choice of the representation space of the
field operators is non-trivial and crucial, and is usually given by 
that of the unperturbed representation in the interaction picture.  
We also use the oscillator variables, adequate for quantum field systems,
rather than $q$ and $p$,
\be
	\left\{
	\begin{array}{c}
	a_\alpha\\
	a_\alpha^\dagger
	\end{array}
	\right\}
	= \frac{1}{\sqrt{2}}\left(\sqrt{\omega} q_\alpha
	\pm i\ve_\alpha\frac{1}{\sqrt{\omega}}p_\alpha \right)\, ,
\ee
Note the factor $\ve_\alpha$ in the definition, and we have no $\ve_\alpha$ in 
the commutation relations, as will be in Eq.~(\ref{eq:CCRa}).
The whole Fock space is spanned by the doubled number eigenstates,
$
	\ket{n_1,n_2} = \ket{n_1}_1 \otimes \ket{n_2}_2 \,,
$
where the introduction of time-reversed state is not necessary
because of the commutation relations without $\ve_\alpha$.
The bra-state is expressed as
$
	\bra{\Psi_b}=\bra {I}= \sum_n \bra{n,n} \,.
$

The formulation of the interaction picture starts with dividing
the total Hamiltonian into the two parts,
$
	H= H_u(t) +H_{I} (t) \,.
$
While $H_I(t)$ represents the interaction effect, $H_u(t)$,
bilinear in the $a$-operators,  defines
the unperturbed representation. The unusual procedure in nonconservative system
is to allow $H_u(t)$ to have the explicit time-dependence and
mixing terms of $(a_1,a_1^\dagger)$ and $(a_2,a_2^\dagger)$, that is,
$H_u(t) \neq  H_{u1}-H_{u2}$ but $H_u(t) = H_{u1}-H_{u2} -Q(t)$
where $Q(t)$ is the mixing Hamiltonian, corresponding to $K$
in the Lagrange formalism in Eq.~(\ref{eq:ClassicalAction}) and brings 
dissipative, nonconservative and irreversible processes. It is called
spontaneous creation of dissipation \cite{Arimitsu}
that no visible
dissipative term is in the total Hamiltonian $H$ but the dissipative
term $Q(t)$ appears in the unperturbed Hamiltonian.

Expect for the mixing term $Q(t)$ in $H_u(t)$,  the fundamental 
properties in the Heisenberg picture should be carried 
to the unperturbed representation.
First we require that the unperturbed Hamiltonian $H_u(t)$ written
in terms of the operators in the interaction picture, denoted by
$a_1(t),a_1^\dagger(t)\, a_2(t),a_2^\dagger(t)$.
\be	\label{eq:H0Tilde}
	\bigl(H_u(t)\bigl)^\sim = -H_u(t) \,,
\ee
with the commutation relations
\be	\label{eq:CCRa}
	[{a^{j}_{\alpha}}(t)\,,\,a^{j',\dagger}_{\alpha'}(t)] =
	 \delta_{\alpha\alpha'}\delta^{jj'}
	\, , \qquad  \mathrm{otherwise} = 0 \, ,
\ee
and the Heisenberg equations
\be	
	i \left\{
	\begin{array}{c}
	{\dot a}_\alpha(t)\\
	{\dot a}^\dagger_\alpha(t)
	\end{array}\right\}
	= \left[  \left\{
	\begin{array}{c}
	{a}_\alpha(t)\\
	{a}^\dagger_\alpha(t)
	\end{array}\right\}
	\,,\, H_u(t) \right]\, .
\ee
The conditions on the bra-state now become
\be	\label{eq:BraEquCond-3}
	\bra{\Psi_b}
	\left\{ \begin{array}{c}
	 a_{1}(t)\\
	a^\dagger_{1}(t)
	\end{array} \right\}
	=\bra{\Psi_b}
	\left\{ \begin{array}{c}
	 a_{2}^\dagger(t)\\
	a_{2}(t)
	\end{array} \right\} \,.
\ee
For consistency, $H_u(t)$ is constrained as
\be	\label{eq:braH0}
	\bra{\Psi_b}H_u(t)=0 \, , \quad \mathrm{i.e.} \quad \bra{\Psi_b}Q(t)=0 \,.
\ee

The unperturbed Hamiltonian $H_u(t)$ and $\ket{\Psi_k}$ uniquely follow from
the requirement that the matrix element of the physical unperturbed number
operator should give the observable expectation of the number, $n(t)$,
\be	\label{eq:n(t)}
	\bra{\Psi_b} a_1^\dagger(t) a_1(t)\ket{\Psi_k}=n(t) \,, 
\ee
with the additional rational conditions. This has already been done in a different
context to derive TFD from the superoperator formalism \cite{NY2013}, 
but the method can be adapted in our present case.
Here we briefly outline the derivations, leaving details to Ref.~\cite{NY2013}.
We assume as in Ref.~\cite{NY2013} that the system is invariant under the following 
global phase transformation
\be
	a_\alpha\, \rightarrow\, e^{i\ve_\alpha \theta} a_\alpha \, ,\qquad
	a^\dagger_\alpha \,\rightarrow \,e^{-i\ve_\alpha \theta} a^\dagger_\alpha
	\, .
\ee
Then Eqs.~(\ref{eq:H0Tilde}), (\ref{eq:braH0}), and (\ref{eq:n(t)}) 
and its time-derivative restrict the form of $H_u(t)$
as
\be
	H_u=\omega \left(a^\dagger_1a_1- a^\dagger_2a_2\right) 
	 \,\,+i \left\{\zeta_1 a_1 a_2+ \zeta_2 a^\dagger_1 a^\dagger_2
	+\zeta_3 \left(a^\dagger_1a_1+ a^\dagger_2a_2 \right)- \zeta_2\right\}\, ,
\ee
where $\omega(t)$ and $\zeta_i(t)$ $(i=1\sim 3)$ are real functions of $t$,
\bea
	\zeta_1(t) &=& {\dot n}(t) +\gamma(t) \,,\nonumber \\
	\zeta_2(t) &=& {\dot n}(t)+ \frac{n(t)}{1+n(t)}\gamma(t)\,,\nonumber \\
	\zeta_3(t) &=&-{\dot n}(t) -\frac{1+2 n(t)}{2(1+n(t))}\gamma(t)
\eea
with an arbitrary real function $\gamma(t)$. If the full causal Green's
function were calculated in the interaction picture, based on the unperturbed 
representation so far, we would see there that the macroscopic 
time-dependent quantity at $t_1$,
say $n(t_1)$, in general affects the microscopic motion in the past, 
at $t_2$ ($t_1> t_2$), which can not be accepted. The macroscopic quantity 
should affect the microscopic motion only in the future, which is called
thermal causality \cite{NY2013}, and holds true if all
 the following equations are satisfied,
\be	\label{eq:BraDota}
	\bra{\Psi_b}{\dot a}_\alpha(t)=\bra{\Psi_b}{\dot a}^\dagger_\alpha(t)
	=0 \, .
\ee  
The necessary and sufficient condition for Eq.~(\ref{eq:BraDota}) is $\gamma(t)=0$,
thus we have
\be 
	H_u= \omega \left(a^\dagger_1a_1- a^\dagger_2a_2\right)
	-i{\dot n} \left(a_2-a^\dagger_1\right)\left(a^\dagger_2-a_1\right) \,.
\ee

We assume that the system approaches to equilibrium after a long time,
i.e. $n(\infty)$ is the equilibrium distribution,
then we have for $\ket{\Psi_k}$ \cite{NY2013},
\be	\label{eq:keta}
	\left\{\left(1+n(t)\right) a_1(t)
	 - n(t) a_2^\dagger(t)\right\}\ket{\Psi_k}=
	\left\{\left(1+n(t)\right) a_2(t)
	 - n(t) a_1^\dagger(t)\right\}\ket{\Psi_k}=0 \,.
\ee
This property with respect to $\ket{\Psi_k}$ is sufficient
to develop the calculations of the full causal Green's function
in the interaction picture. Because of $\bra{\Psi_b} H_I(t)=0$ and 
therefore $\bra{\Psi_b} U^{-1}(t_f,t_i)=\bra{\Psi_b}$, 
we can show that
\be	\label{eq:Dysonformula}
	\bra{\Psi_b}{\rm T}\left[A_H(t)B_H(t) \cdots\right]\ket{\Psi_k}
	=\bra{\Psi_b}{\rm T}\left[U(t_f,t_i) A(t)B(t) \cdots\right]\ket{\Psi_k}
	\, ,
\ee
with
$
	i{\partial}U(t,t_i)/{\partial t} =H_I(t) U(t,t_i)\,, U(t_i,t_i)=I \,.
$
Equations (\ref{eq:BraEquCond-3}) and (\ref{eq:keta}) allow us to 
apply the Wick theorem in the Dyson formula in Eq.~(\ref{eq:Dysonformula}), 
so that the Feynman diagram method can be used. What has been derived is
identical with the nonequilibrium TFD \cite{AIP, NY2011}.

The doubling of degrees of freedom and dissipation of quantum system have also 
been discussed in a way \cite{CRV,BVJ}, different from the present paper.

\section{Reservoir model}
We apply the above formulation to the reservoir model \cite{Arimitsu, Advances}
for clarity, 
whose total Hamiltonian in the case of doubled degrees of freedom is given by $H=H_1-H_2$ with 
\be
	H_\alpha = \omega_0 a_\alpha^\dagger a_\alpha +  \sum_k\bigl[ \Omega_{0k} 
	  A_{\alpha k}^\dagger A_{\alpha k} + 
       g_k (a_\alpha^\dagger A_{\alpha k} + A_{\alpha k}^\dagger a_\alpha) \bigl]\,.
\ee
Here $a$ and $A_k$ represent the degrees of freedom of the system 
under consideration and those of the reservoir, respectively, and we
set $\hbar=1$. Note that this Hamiltonian is invariant under the global phase transformation.
The coupling constant $g_k$ is considered to be small and of order of $1/\sqrt{N}$ where $N$ is 
the total number of $A_k$ and is taken to be infinite at the final stage.
The unperturbed and interaction Hamiltonians are chosen to be
\bea
	H_u(t)\!&=& \!\sum_\alpha \ve_\alpha\Bigl[
	\omega(t) a^\dagger_\alpha a_\alpha 
	+\sum_k \Omega_k A^\dagger_{\alpha k}A_{\alpha k} \Bigr]-Q(t)\,,\\
 	H_I(t)\!&=& \!\sum_\alpha\ve_\alpha \Bigl[
	\sum_k g_k \bigl( a_\alpha^\dagger A_{\alpha k} + A_{\alpha k}^\dagger a_\alpha \bigr)
	-\delta\omega(t)a^\dagger_\alpha a_\alpha\Bigr] +Q(t) \,,
\eea
with $Q(t)=i{\dot n}(t)(a_2- a^\dagger_1)(a^\dagger_2-a_1)$,
where the time dependence of operators is suppressed and the renormalized energy
of the system is given by $\omega(t)= \omega_0 +\delta \omega(t)$.  Because of 
the weak coupling, the self-energy corrections for each $A_k$ vanish and therefore
$\delta \Omega_k(t)={\dot N}_k(t)=0$ and $\Omega_k=\Omega_{0k}$.
The renormalization conditions on the self-energy, which we have
proposed in Ref.~\cite{NY2013} as a generalization of the on-shell
 renormalization, derive the quantum transport equation and determine
the renormalized energy for the system as
\bea
	\label{eq:RM-dotn}
	&&{\dot n}(t) \!= -2\mathrm{Re} \sum_k g^2_k \!\int^{t}_{t_i} \!\!ds \, 
	\{n(s)-N_k \}
	e^{i\int_{s}^{t} \!ds'[\omega(s')-\Omega_k]} \,,\\
	\label{eq:RM-domega}
	&&\delta\omega(t)\! = \mathrm{Im} \sum_k g^2_k \!\int^{t}_{t_i} \!\!ds \, 
	e^{i\int_{s}^{t} \!ds'[\omega(s')-\Omega_k]} \,.
\eea
When we put $\bar{g}^2=Ng^2_k \sim O(1)$, $\Omega_k=\bar{\Omega}+k\delta$, $\delta\sim O(1/N)$ and 
take the limit $N\to \infty$, the sum is replaced by the integral as 
$\sum_k g_k^2 \cdots = \bar{g}^2/\Delta \int\!d\Omega\cdots\,,$ with the band width $\Delta=N\delta$.
In the long-time limit, namely $t-t_i\to\infty$, we find a consistent solution for Eq.~(\ref{eq:RM-domega}) 
with time-independent $\delta\omega$ and $\omega$, 
\be
	\delta\omega = \frac{\bar{g}^2}{\Delta} \int\! d\Omega \;\mathcal{P}\frac{1}{\omega-\Omega} \,.
\ee
Furthermore, in such a case the thermal change is so slow that $n(s)$ can be replaced with
$n(t)$ in Eq.~(\ref{eq:RM-dotn}) (the Markovian ansatz), which is reduced to
\be
	{\dot n}(t)= -2 \kappa \left\{n(t)-N_\omega \right\}\,,
\ee
with $\kappa= {\pi {\bar g}^2}/{\Delta}\,$,
and shows that $n(t)$ approaches to equilibrium at $t\to\infty$ with the relaxation time $1/2\kappa$.

\section{Summary}
In summary, the Hamilton formalism of classical nonconservative systems
by Galley can be extended to quantum field while the physical limit and equality conditions are
realized as the requirements (a) and (b). Then the doubling of degrees of freedom comes from 
nonconservative nature, but not necessarily from quantum nature. Moreover, the
formalism for quantum field, formulated in the interaction picture, is the nonequilibrium TFD
 \cite{AIP, NY2011, NY2013} under the additional assumptions of the thermal causality and the relaxation to equilibrium
at $t=\infty$.  For illustration, the formalism is applied to the reservoir model,
where the renormalization condition derives the kinetic equation. 

\section*{Acknowledgments}
The authors thank Prof.~S.~Abe at Mie University for communicating us 
Galley's paper. This work is partly supported by Grant-in-Aid for Scientific Research (C) (No. 25400410) from the Japan Society for the Promotion of Science, Japan and by Waseda University Grant for Special Research Projects (Project No. 2013A-876). 


\end{document}